\documentclass[manuscript]{aastex}

\newcommand{\GALEX}{{\it GALEX}}

\newcommand{\HST}{{\it HST}}

\newcommand{\Teff}{T_{\rm eff}}

\newcommand{\kms}{{\>\rm km\>s^{-1}}}

\begin{document}

\title{A Luminous Yellow Post-AGB Star in the Galactic Globular
Cluster~M79\altaffilmark{1}}

\author{
Howard E. Bond,\altaffilmark{2,3,4}
Robin Ciardullo,\altaffilmark{2}
and
Michael H. Siegel\altaffilmark{2}
}

\altaffiltext{1}
{Based in part on observations with the 1.3- and 1.5-m telescopes operated by
the SMARTS Consortium at Cerro Tololo Interamerican Observatory.}

\altaffiltext{2}
{Department of Astronomy \& Astrophysics, Pennsylvania State
University, University Park, PA 16802, USA; heb11@psu.edu}

\altaffiltext{3}
{Space Telescope Science Institute, 
3700 San Martin Dr.,
Baltimore, MD 21218, USA}

\altaffiltext{4}
{Visiting astronomer, Cerro Tololo Inter-American Observatory, National Optical
Astronomy Observatory, which is operated by the Association of Universities for
Research in Astronomy (AURA) under a cooperative agreement with the National
Science Foundation.  
}

\begin{abstract}

We report discovery of a luminous F-type post-asymptotic-giant-branch (PAGB)
star in the Galactic globular cluster (GC) M79 (NGC~1904). At visual apparent
and absolute magnitudes of $V=12.20$ and $M_V=-3.46$, this ``yellow'' PAGB star
is by a small margin the visually brightest star known in any GC\null. It was
identified using CCD observations in the {\it uBVI\/} photometric system, which
is optimized to detect stars with large Balmer discontinuities, indicative of
very low surface gravities. Follow-up observations with the SMARTS 1.3- and
1.5-m telescopes show that the star is not variable in light or radial velocity,
and that its velocity is consistent with cluster membership. Near- and
mid-infrared observations with 2MASS and {\it WISE\/} show no evidence for
circumstellar dust. We argue that a sharp upper limit to the luminosity function
exists for yellow PAGB stars in old populations, making them excellent
candidates for Population~II standard candles, which are four magnitudes
brighter than RR~Lyrae variables. Their luminosities are consistent with the
stars being in a PAGB evolutionary phase, with core masses of
$\sim\!0.53\,M_\odot$.

We also detected four very hot stars lying above the horizontal branch
(``AGB-manqu\'e'' stars); along with the PAGB star, they are the brightest
objects in M79 in the near ultraviolet. In an Appendix, we give periods and
light curves for five variables in M79: three RR~Lyrae stars, a Type~II Cepheid,
and a semiregular variable.

\end{abstract}


% \clearpage

\section{Introduction}

The most luminous stars in old populations are expected to be those that have
departed from the top of the asymptotic giant branch (AGB) and are evolving to
the left in the Hertzsprung-Russell diagram (HRD) at nearly constant luminosity.
Around spectral type F, bolometric corrections are smallest, making these
``yellow'' post-AGB (PAGB) stars the visually brightest members of globular
clusters (GCs). The tenth-magnitude F-type star ROA~24 in $\omega$~Centauri
(also cataloged as HD~116745 and ``Fehrenbach's star'') is the only star in any
GC bright enough to be listed in the {\it Henry Draper Catalogue\/} (Harding
1965; Dickens \& Powell 1973; Gonzalez \& Wallerstein 1992). It has often been
considered to be a PAGB star---but see our discussion below in \S5.3. The
Galactic GC NGC~5986 contains two yellow PAGB stars (Alves, Bond, \& Onken
2001), likewise the brightest members of their cluster. In the HRD, stars of
this rare type lie some four magnitudes above the horizontal branch (HB) and
RR~Lyrae stars.

Analogs of these objects are known in the field. Two examples are
BD~+39$^\circ$4926 (Slettebak, Bahner, \& Stock 1961) and HD~46703 (Bond 1970;
Luck \& Bond 1984). Both are long-period spectroscopic binaries, with orbital
periods of 775~days for BD~+39$^\circ$4926 (Kodaira, Greenstein, \& Oke 1970) 
and 600~days for HD~46703 (Hrivnak et al.\ 2008). There is no published evidence
that BD~+39$^\circ$4926 has a variable brightness. However,  the cooler HD~46703
lies in a high-luminosity extension of the RR~Lyrae pulsational instability
strip; it is a semiregular variable with a range of $\sim$0.3~mag and a main
period of about 29~days (Bond et al.\ 1984; Hrivnak et al.\ 2008; Aikawa 2010).

It has been suggested (Bond 1997a,b; see also Vickers et al.\ 2015) that F- and
A-type PAGB stars should be useful Population~II standard candles, because on
theoretical grounds there should be a sharp upper limit to their luminosity
function. Moreover, they are easy to detect using broadband photometry on small
telescopes, as their low surface gravities give them extremely large Balmer
jumps.  In the late 1990's, H.E.B. and collaborators began to develop a
photometric system optimized for the efficient discovery of such stars. This
system combines the $u$ filter of Thuan \& Gunn (1976), whose bandpass lies
almost entirely shortward of the Balmer jump, with the standard broadband
Johnson-Kron-Cousins {\it BVI\/} filters. The design principles of this {\it
uBVI\/} system were described by Bond (2005, hereafter Paper~I). A network of
standard stars was established by Siegel \& Bond (2005, hereafter Paper~II),
based on extensive observations with CCD cameras at 0.9- and 1.5-m telescopes at
Kitt Peak National Observatory (KPNO) and Cerro Tololo Interamerican Observatory
(CTIO).

% Unlike Cepheids, PAGB stars do not require a long sequence of
% observations for their discovery.

In this paper we report our discovery, using {\it uBVI\/} photometry, of a
bright PAGB star in the GC M79 (NGC~1904). We then present follow-up photometric
and spectroscopic monitoring aimed at placing limits on the variability and
binarity of the star. We close with a negative search for circumstellar dust, an
argument for the case that Population~II PAGB stars are indeed promising
standard candles, and a discussion of their evolutionary status. An appendix
presents photometry of several known variable stars in~M79.

\section{{\em uBVI\/} Observations and Discovery of the PAGB Star}

As part of our {\it uBVI\/} survey of the Galactic GCs, M79 was imaged on 1997
November~7 with the CTIO 0.9-m telescope and a Tek $2048\times2048$ CCD
detector, giving a field of view of $13\farcm5\times13\farcm5$. Exposure times
in {\it uBVI\/} were 400, 60, 45, and 60~s, respectively. These are relatively
shallow frames, aimed at detecting luminous PAGB stars without saturating their
images.

The frames were bias-subtracted and flat-fielded using the {\tt CCDPROC} task in
IRAF\footnote{IRAF is distributed by the National Optical Astronomy Observatory,
which is operated by the Association of Universities for Research in Astronomy
(AURA) under a cooperative agreement with the National Science Foundation.}, and
instrumental stellar magnitudes were measured with the DAOPHOT tasks {\tt
ALLSTAR} and {\tt DAOGROW} (Stetson 1990).  After correction for atmospheric
extinction, the photometry was calibrated to the {\it uBVI\/} system using
observations on the same night of standard-star fields from Paper~II\null.

% 9/25/15: conversion from Siegel RA/Dec to x,y in obj1303.fits :
% 
% PAGB star is at RA/Dec;  5.40288 -24.48906  x,y = 999.90  692.09
% two other bright stars:  5.39706 -24.54469       1711.83 1196.57 
%                          5.40254 -24.57114       1039.58 1429.85
% RA table: 
% 5.40288  999.90
% 5.39706 1711.83
% 5.40254 1039.58
% 
% Dec table:
% 
% -24.48906  692.09
% -24.54469 1196.57
% -24.57114 1429.85
% 
% Fit gives: x = -122478.1282*ra + 662733.5773
% 	   y = -9000.663443*dec - 219724.7775
% 
% So the AGB-M stars are here:
% 
% 5.40374  -24.53908	893.616 1143.222  nice			ID 2038
% 5.40246  -24.53336     1050.388 1091.738  nice	           4119
% 5.403525 -24.51205	919.949  899.934  in very close pair     2370= UIT 2
% 5.403313 -24.53892	945.914 1141.782  very faint             2693= UIT 1
% 
% Mike emailed that the last 2 were added back in based on UVOT observations.

We show the resulting color-magnitude diagrams (CMDs) of M79 in $V$ vs.\ $B-V$
in Figure~1a, and in $V$ vs.\ $V-I$ in Figure~1b. To reduce field-star
contamination (which is already quite small), we plot only stars lying within
$4\farcm8$ from the cluster center. Because of source confusion, all of the
plotted stars are at least $0\farcm5$ from the center. The data have not been
corrected for the cluster's very small reddening of $E(B-V)=0.01$ (Harris
1996\footnote{Catalog of Parameters for Milky-Way Globular Clusters, 2010
edition at \tt http://www.physics.mcmaster.ca/$^\sim$harris/mwgc.dat}). Harris
gives an M79 distance of 12.9~kpc and a metallicity of $\rm[Fe/H]=-1.60$. 

CMDs for M79, generally reaching considerably deeper than ours, have been
published previously by a number of authors, including Ferraro et al.\ (1992),
Alcaino et al.\ (1994), Kravtsov et al.\ (1997), Rosenberg et al.\ (2000),
Lanzoni et al.\ (2007), Dalessandro et al.\ (2012), and Kopacki (2015). These
studies used ground-based CCD data and, in the cases of Dalessandro et al.\ and
Lanzoni et al., space-based {\it Hubble Space Telescope\/} (\HST\/) and {\it
Galaxy Evolution Explorer\/} (\GALEX\/) observations. All of these authors have
shown that the cluster has an extremely blue HB with a long blue tail, and a
sparsely populated AGB---a CMD that is remarkably similar to that of the
better-known M13. Our {\it BVI\/} data plotted in Figures~1a and 1b are in
complete agreement with these findings. 

In Figure~1c we plot the $(u-B)-(B-V)$ color difference vs.\ $V-I$\null. The
color difference is an analog in the {\it uBVI\/} system of the $c_1$ index in
the Str\"omgren system; it measures the size of the Balmer jump, while being
relatively insensitive to reddening and metallicity (see Paper~I for details).
The plot shows only stars brighter than $V=17.75$, and individual stars with
errors in either coordinate greater than $\pm$0.06~mag have been deleted. The
distribution of objects peaking on the left-hand side of this figure contains
the HB stars, while the sequence of objects on the right-hand side are the
subgiants and red giants, plus a small number of AGB stars. The objects clumped
around $V-I\simeq0.65$ are a few foreground field stars.

Five stars stand out in the CMDs and color-color diagrams. Most conspicuous is
the very bright PAGB candidate, marked with red filled circles in the three
plots; this star is prominent in the $(u-B)-(B-V)$ vs.\ $V-I$ diagram because of
its unusually large Balmer-jump index. Stars with such colors are extremely
rare, leaving little doubt that the object is a PAGB star belonging to the
cluster. There are also four hot, blue stars lying above the HB, marked with
blue filled circles in all three plots in Figure~1.  Such objects are usually
interpreted as stars evolving off the blue HB, destined to become white dwarfs
without ascending the AGB---the so-called AGB-manqu\'e (AGB-M) stars. Table~1
lists properties of these five stars. A finding chart is presented in Figure~2.
The two optically faintest AGB-M objects are extremely hot stars, which are very
bright in the far- and near-ultraviolet (FUV and NUV, respectively). They had
been noted by Hill et al.\ (1992) in their {\it Astro-1\/} Spacelab Ultraviolet
Imaging Telescope (UIT) observations of M79, and were designated UIT~1 and
UIT~2.

% 10/1/15: To make chart, first have to flip the u image in the y coord:
% 	imcopy obj1301.fits[*,-*] obj1301_flip
% 	disp obj1301_flip[0583:1357,0663:1437] 1 zr- zs-
% so the image displayed is 775x775 pixels * .384"/pix = 5'.0 x 5'.0
% Put on circles with:
% 	tvmark 1 labels.dat mark="circle" radii=11 color=205 label- 
% 	tvmark 1 labels2.dat mark="circle" radii=11 color=205 label- 
% the labels were put on the u-band image with:
% tvmark 1 labels.dat mark="none" color=205 label+ txsize=2 nxoff=15 nyoff=-10
% tvmark 1 labels2.dat mark="none" color=205 label+ txsize=2 nxoff=-85 nyoff=-10

At an absolute magnitude of $M_V=-3.46$ (see below), the M79 PAGB star is by far
the visually brightest member of its cluster, and in fact marginally the
visually absolutely brightest star known in any GC\null. To our knowledge, in
spite of the many investigations of M79 referenced above, the star had not been
recognized previously as belonging to the cluster. In most studies it goes
unmentioned, and has apparently been dismissed as a foreground star; actually,
in typical deep CCD exposures, its image would be saturated. The object was,
however, designated as star ``H'' in a photographic study of M79 by Stetson \&
Harris (1977), who used it as one of their calibration stars; they gave
photoelectric measurements of $V=12.21$, $B-V=0.30$, and $U-B=0.24$. Hill et
al.\ (1992) noted its brightness in the NUV---it is the brightest star in their
NUV frame---but considered it to be ``an early F-type field star'' (it being
very faint in their FUV image).

Our discovery of the PAGB star in M79 was communicated privately to {\c S}ahin
\& Lambert (2009, hereafter SL09), who obtained high-resolution spectra at
McDonald Observatory and carried out an abundance analysis. SL09 reported a very
low iron content and a radial velocity (RV) of $+211\pm5\kms$, consistent with
the cluster's RV of $+205.8\pm0.4\kms$ and velocity dispersion of
$5.3\pm0.4\kms$ (Harris 1996). Both the low metallicity and the velocity clearly
establish the star's membership in the GC\null. Their spectroscopic analysis
yielded an effective temperature and surface gravity of $\Teff=6300$~K and $\log
g=0.8$. The very low $\log g$ implied by our {\it uBVI\/} photometry was thus
verified. The stellar parameters of the M79 PAGB star are very similar to those
of the field PAGB star HD~46703 ($\Teff=6000$~K, $\log g=0.4$; Luck \& Bond
1984). An interesting result of the SL09 analysis was that its iron abundance,
$\rm[Fe/H]=-2.0$, is about 0.6~dex lower than that of the cluster's red giants.
A similar effect has been seen in field PAGB stars (e.g., Bond 1991), and
discussed in terms of depletion of metals onto grains (e.g., Hrivnak et al.\
2008 and references therein), but for field stars the progenitor's original
metallicity is considerably less certain.

M79 has been imaged more recently in the NUV with the Ultraviolet Optical
Telescope (UVOT) onboard the {\it Swift\/} gamma-ray-burst satellite (Siegel et
al.\ 2014). In the UVOT's {\it uvw2\/} bandpass at $\sim$1928~\AA, the PAGB star
and the four AGB-M stars of our Table~1 are the five brightest stars in M79
(Siegel et al., their Figure~7). 

\section{Photometric Monitoring}

As noted above, a subset of F-type PAGB stars like HD~46703  are pulsating
variables, lying in an extension of the RR~Lyrae instability strip. The M79 PAGB
star has an effective temperature similar to that of HD~46703 (although both the
spectroscopic analyses and optical photometry do imply it to be slightly
hotter). We have carried out a program of photometric monitoring to determine
whether the M79 star is likewise a pulsating variable.

The observations, made by Chilean service personnel, used the 1.3-m SMARTS
Consortium\footnote{SMARTS is the Small \& Moderate Aperture Research Telescope
System; {\tt http://www.astro.yale.edu/smarts}} telescope at CTIO and its
ANDICAM CCD camera. Data were obtained on the M79 PAGB star and surrounding
field on 224 nights between 2007 February 20 and 2011 May~8. A typical cadence
was every five days, chosen so that a pulsation period of around one month would
be easily detected. On each night, two exposures of 30~s each were obtained in
$B$ and in $V$\null. The usable field of view of these images is about
$5\farcm6\times5\farcm6$.

The ANDICAM frames were bias-subtracted and flat-fielded in the SMARTS pipeline
at Yale University. We then carried out photometry on the reduced images. The
point-spread function (PSF) of each frame was determined with the DAOPHOT {\tt
ALLSTAR} routine, using 18 bright, isolated stars in the field. The central
portion of this PSF (a region slightly larger than the FWHM) was then fit to
both the program object and several nearby targets chosen as photometric
comparison stars, in order to determine differential magnitudes. Instrumental
magnitudes determined in this way were generally precise at a level of
$\sim$0.005~mag for the PAGB star, and $\sim$0.007~mag for typical fainter
comparison stars, except for a few frames affected by thick clouds or poor
seeing.

For the final analysis, we determined differential magnitudes between the PAGB
star and the sum of intensities of eight nearby, constant-brightness comparison
stars. Since the PAGB star is bluer than all of the comparison stars (which are
a combination of field stars and bright red giants in the cluster), we removed a
small mean trend with airmass in the $B$ magnitudes (there was no airmass
dependence detected for the $V$ magnitudes).

In Figure~3 we plot the differential magnitudes in $B$ and $V$, with arbitrary
zero points. We see no evidence for variability, certainly not at the level of
the $\sim$0.3~mag range of HD~46703. There is marginal evidence for slow,
low-amplitude variations, particularly during 2009, but these may well be subtle
seasonal variations of an instrumental origin. 

We also took advantage of this collection of frames for a study of several known
variable stars in M79; see Appendix~A.

\section{Radial-Velocity Monitoring}

As noted in the Introduction, several of the known metal-poor A- and F-type PAGB
stars in the field are known to be long-period spectroscopic binaries. To
investigate whether the M79 PAGB star is also a binary, we monitored its RV on
67 nights over the interval from 2007 February~21 to 2012 January~29. The
queue-scheduled spectroscopic observations were made by Chilean service
observers with the SMARTS 1.5-m telescope at CTIO, using the RC-focus
spectrograph equipped with a CCD camera. Two different setups were employed with
grating number 47: setting ``47/II,'' covering 3878--4552~\AA, was used
primarily during the first half of the observing interval; during the second
half, setting ``47/IIb,'' covering 4070--4744~\AA, was used for most of the
observations. Both setups yielded a spectral resolution of 1.6~\AA\null.
Exposure times each night were generally $3\times300$~s, and short exposures of
a HeAr calibration lamp were taken before and after each set of stellar
observations.

The CCD images were bias-subtracted and flat-fielded, combined for cosmic-ray
removal, and then the stellar spectrum was extracted and wavelength-calibrated,
all using standard IRAF routines. In Figure~4, we show a spectrum created by
combining all of the data, and normalizing to a flat continuum. The spectrum is
dominated by sharp, strong lines of the Balmer series and \ion{Ca}{2} H and
K\null. All other lines are relatively quite weak. 

On about half of the nights, an RV standard star was observed with the same
grating setup. The standard stars we used were 6~Ceti (2 spectra with 47/II, 1
with 47/IIb), 5~Serpentis (2 spectra with 47/IIb), 10~Tauri (4 spectra with
47/II, 15 with 47/IIb), and $\beta$~Virginis (2 spectra with 47/II, 6 with
47/IIb). We adopted the absolute RVs for these stars given by Stefanik, Latham,
\& Torres (1999).

We used the IRAF task {\tt fxcor}, which determines RVs (including corrections
to the heliocentric frame) through cross-correlation of the program spectrum
with a spectrum of a star of known RV\null. We employed a two-stage process.
First, each individual spectrum of the M79 star was cross-correlated with each
of the standard-star spectra taken with the same setup, and for each PAGB
spectrum we adopted the mean of these determinations. Because the \ion{Ca}{2} H
and K features in the standard stars are much stronger and broader than in the
PAGB star, and because the metallic lines are so weak in the PAGB star, we only
used small 40~\AA-wide wavelength windows around H$\delta$ and H$\gamma$ for
this first pass. The mean RV of the PAGB star was found to be $+218.5\kms$ for
the 47/II observations, and $+207.0\kms$ for 47/IIb, with standard deviations of
5.0 and $7.5\kms$, respectively. 

The spectra of the PAGB star and the standard stars, which are F-type subgiants
and dwarfs with much higher metallicities and gravities, are not very well
matched. Therefore, in a second pass, we cross-correlated each individual PAGB
spectrum with the ensemble of all PAGB spectra taken with the same setup,
setting the standard velocity of each spectrum in the ensemble  to the mean
velocity from the first pass, $+212.9\kms$. In the second pass, the entire
spectral wavelength range was used in the correlations. This resulted in mean
RVs of $+213.1\kms$ for the 47/II spectra, and $+213.4\kms$ for the 47/IIb
spectra, with standard deviations of 4.7 and $5.5\kms$, respectively.

Figure~5 plots our final velocities versus time. There is no evidence for
variations in excess of the errors. The peak-to-peak RV variations for
BD~+39$^\circ$4926 and HD~46703 are $\sim$30 and  $\sim\!34\kms$, respectively
(Kodaira, Greenstein, \& Oke 1970; Hrivnak et al.\ 2008). Velocity variations of
this size are definitely ruled out for the M79 PAGB star, but, given the modest
precision of our measurements, we cannot exclude a binary with a low RV
amplitude, or one viewed close to pole-on.

\section{Discussion}

\subsection{No Evidence for Circumstellar Dust}

% Data for the SED
% 
% Filt	w.l.		magnitude	Flambda
% 
% GALEXfuv20.56
% GALEXnuv-999
% 
% UVOT, estimated from the plots in Siegel et al, and zero-points (see my
% constants.txt file for details):
% 
% uvw2	0.1928		15.0		0.555e-14
% uvw1	0.26		13.82		1.122e-14
% 
% u			13.803
% B	0.44		12.480		6.31e-14
% V	0.55		12.203		4.69e-14
% I	0.79		11.744		2.35e-14
% 
% J	1.235		11.274		0.97e-14
% H	1.662		11.069		0.423e-14
% Ks	2.159		11.028		0.166e-14
% 
% w1	3.35		10.857		0.0376e-14
% w2	4.60		10.910		0.0105e-14
% w3	11.56		10.854		0.000324e-14
% w4	22.09		 9.203 (undetected, as far as I can tell--
% 				has a negative SNR; might be an upper limit?)
%                       so the UL is <  0.000107e-14
% 
% 
% 2mass J05241036-2429206

Substantial circumstellar (CS) dust is a well-known feature of PAGB stars
belonging to Galactic-disk populations (see the review by van Winckel 2003, and
references therein). To search for CS dust around the M79 PAGB star, we
determined its spectral-energy distribution (SED) by combining our optical
photometry (Table~1) with the {\it Swift\/} UVOT measurements by Siegel et al.\
(2014), and near- and mid-IR photometry from 2MASS (Skrutskie et al.\ 2006) and
{\it WISE\/} (Wright et al.\ 2010). These data points are plotted in Figure~6.
Also shown is a blackbody spectrum for $\Teff= 6300$~K, which is the effective
temperature of the M79 star determined in the spectroscopic analysis by SL09.

As the figure illustrates, the blackbody curve fits the SED extremely well.
There is no evidence for CS dust out to a wavelength of $12\,\rm\mu m$, and the
star was not detected by {\it WISE\/} at $22\,\rm\mu m$. The lack of dust may
plausibly be explained in part by the star's low metallicity, making dust
formation difficult in the first place; however, the photospheric iron
depletion, mentioned in \S2, may require that there was some dust formation in
the past. In any case, the long post-AGB evolutionary timescales for low-mass
remnants in old stellar populations may allow sufficient time for any CS dust
that was formed to have dissipated. Note also that near- or mid-IR excesses have
not been detected in the field analogs BD~+39$^\circ$4926 and HD~46703.

A further conclusion is that the M79 PAGB star will not produce a planetary
nebula (PN)\null. It has been argued that the few PNe known in Milky Way and
Local Group GCs are descended from binary-star interactions, and not from single
stars (e.g., Jacoby et al.\ 2013; Bond 2015; and references therein).  

\subsection{F- and A-Type PAGB Stars as Standard Candles}

As mentioned in \S1, it has been proposed that yellow PAGB stars in old
populations are promising standard candles. In Table~2, we assemble $V$-band
photometry for four such candidates known in GCs: ROA~24 in $\omega$~Cen, the
two in NGC~5986, and the M79 star discussed in this paper. We calculated visual
absolute magnitudes, using the visual apparent distance moduli from the sources
cited in the table. 

ROA~24 deserves some discussion. It has traditionally been considered to be a
PAGB star, along with the visually second-brightest star in $\omega$~Cen, the
RV~Tauri-type variable V1 (e.g., Gonzalez \& Wallerstein 1994; van Loon et al.\
2007; McDonald et al.\ 2011). However, the bolometric luminosities of both stars
are below that of the AGB tip in $\omega$~Cen (for example, McDonald et al.\
2009, their Fig.~3; McDonald et al.\ 2011, their Fig.~1). This suggests that
ROA~24 and V1 may more properly be regarded as post-early-AGB stars (e.g.,
Dorman et al.\ 1993, their Fig.~1), rather than stars evolving off the tip of
the AGB\null. In our Table~2, the absolute magnitude of ROA~24 is indeed
$\sim$0.2--0.3~mag fainter than for the other three PAGB candidates.

Setting aside ROA~24, we find a mean visual absolute magnitude for the remaining
three stars in Table~2 of $M_{V,0}=-3.38$, with a standard deviation
of 0.09~mag, and an error of the mean of only 0.05~mag. 

This scatter is considerably smaller than the cosmic scatter in the
period-luminosity relation for Cepheid variables.  There is no evidence for
variability or circumstellar dust for any of the stars, strengthening the case
that they are well-behaved candidates for standard candles. They are
intrinsically four magnitudes brighter than RR~Lyrae stars, and as bright as
Cepheids with periods of about 5.3~days. Moreover, unlike RR~Lyrae stars and
Cepheids, PAGB stars do not require a long series of photometric observations
for their discovery: the total exposure time, with a 0.9-m telescope, needed to
discover the M79 star was less than 10~minutes. They are expected to be present
in old populations that do not contain Cepheids, including elliptical galaxies
and dwarf spheroidals. And these objects should exist in halos of spiral
galaxies, where there are fewer complications due to interstellar extinction
than encountered for Population~I Cepheids in the disks. 

A practical difficulty is the existence of fainter, but still very low-gravity,
stars like ROA~24 and V1 in $\omega$~Cen. For PAGB stars to be a robust distance
indicator, it will be necessary to sample a large enough population to assure
that the upper edge of the luminosity function has been detected.

In subsequent papers, we will report searches for additional PAGB stars in GCs
and in the halo of M31. In the next few years, we expect that parallaxes from
{\it Gaia\/} will provide an excellent zero-point calibration, based on PAGB
stars in the field, as well as in Galactic~GCs.

\subsection{Evolutionary Status}

For a mean $M_{V}=-3.38$, and neglecting the small difference in bolometric
corrections between them and the Sun ($M_V=+4.86$), the PAGB candidates in
NGC~5986 and M79 have an average luminosity of $\log L/L_\odot\simeq3.30$. Using
the theoretical core-mass\slash luminosity relation for PAGB stars of
Vassiliadis \& Wood (1994), $L/L_\odot = 56694 (M/M_\odot - 0.5)$, we find a
core mass of $\sim\!0.53\,M_\odot$. This is in excellent agreement with the
masses inferred for white dwarfs in GCs (e.g., Kalirai et al.\ 2009 found
$0.53\pm0.01\,M_\odot$ for the white dwarfs in four Galactic GCs). Thus the case
is strengthened that these three stars are in their final PAGB evolutionary
stages, just before arriving at the top of the white-dwarf cooling sequence.
However, we cannot entirely rule out more complicated scenarios such as those
involving binary interactions. Complete surveys for these objects in the
Galactic GC system, and in the halos of nearby spirals, would provide
information on the lifetimes of the F- and A-type PAGB stars, and thus shed more
light on their evolutionary status.

% Study of variables:
% http://adsabs.harvard.edu/abs/2012A%26A...548A..92K

% HUT spectrum, Hot stars:
% http://adsabs.harvard.edu/abs/1999A%26A...345..109V

% Horiz br/HUT:
% http://adsabs.harvard.edu/abs/1996AJ....111.1936D

\acknowledgments

We thank the STScI Director's Discretionary Research Fund for supporting our
participation in the SMARTS consortium, Fred Walter for scheduling the 1.5-m 
queue observations, and Phil Massey for advice on the use of {\tt fxcor}. We
especially appreciate the excellent work of the CTIO/SMARTS service observers
who obtained the direct images and spectra during many long clear Tololo nights:
% 1.5m data:
% additional from 1.3m
Claudio Aguilera,
Edgardo Cosgrove,
Juan Espinoza,
Manuel Hern\'andez, 
Rodrigo Hern\'andez,
Alberto Miranda,
Alberto Pasten,
Jacqueline Seron,
John Subasavage,
Joselino Vasquez,
and Jos\'e Vel\'asquez.
This publication makes use of data products from the Two Micron All Sky Survey,
which is a joint project of the University of Massachusetts and the Infrared
Processing and Analysis Center/California Institute of Technology, funded by the
National Aeronautics and Space Administration and the National Science
Foundation.
It also makes use of data products from the Wide-field Infrared Survey
Explorer, which is a joint project of the University of California, Los Angeles,
and the Jet Propulsion Laboratory/California Institute of Technology, funded by
the National Aeronautics and Space Administration.

{\it Facilities:} 
\facility{CTIO:0.9-m, CTIO:1.3m, CTIO:1.5m}

\appendix

\section{Variable Stars in M79}

The population of variable stars in M79 has been studied by several authors,
including Rosino (1952), Amigo et al.\ (2011), Kains et al.\ (2012), and Kopacki
(2015). In our photometric monitoring with the SMARTS 1.3-m telescope, described
in \S3, the relatively small field was centered on the PAGB star and thus
covered only a portion of the cluster. Nevertheless, one candidate Population~II
Cepheid, a red giant of uncertain variable type, and three RR~Lyrae variables
are contained in our field, so we were able to obtain data on them over five
observing seasons between 2007 and 2011. Table~3 lists these variables and the
elements (zero phase and period) used to prepare the phased light curves given
below. The designations are those of the online catalog of variables in GCs
maintained by C.~Clement\footnote{Version of 2013 July at \tt
http://www.astro.utoronto.ca/$^\sim$cclement/cat/listngc.html}. We did not
attempt photometry of several low-amplitude variable red giants and SX~Phe stars
identified by Kopacki (2015), nor of variables in the crowded central regions of
the cluster.

Figure~7 shows the nominal locations of the five variables in the cluster's
CMD\null. The photometric measures were reduced as described in \S3, i.e., we
determined differential magnitudes between each variable and the intensity sum
of eight constant stars in the field. For purposes of plotting light curves, the
magnitude zero points were set to arbitrary values. Details of the results are
as follows.

{\bf V6:} This RRc variable was discovered by Rosino (1952), and
extensive data have been obtained by Amigo et al.\ (2011), Kains et al.\ (2012),
and Kopacki (2015). The modern authors have commented that the period appears to
be variable, and we have confirmed this: it proved impossible to fit a single
period to all of our data. By breaking our data into five individual annual
seasons, and finding periods with a Lafler-Kinman (1965) periodogram, we did
find that each season's data could be fitted reasonably well with a single
period. However, the period changes from season to season. By making slight
adjustments in the epochs of zero phase, it was possible to superpose all of the
seasonal light curves. Table~4 summarizes the elements used for this
superposition. Figure~8 shows the resulting light curves, with color coding to
indicate the individual seasons. The light curve shape is remarkably constant
over our observing interval; only the period and zero-point appear to change. We
investigated whether this might be due to a light-travel-time effect in a binary
system, but could not find convincing evidence for such an interpretation. The
variable period appears instead to be a Blazhko effect.

{\bf V7 and V8:} These two variables were discovered by Chu (1974)\footnote{We
thank Profs.\ You-Hua Chu (no relation) and Yang Chen for obtaining a digital
scan of this paper and providing an English summary.}, but little was known
about them until quite recently. V7 was classified as a W~Virginis-type
Population~II Cepheid by Kopacki (2015) with a period of 13.985~days, although
this was based on only a few cycles of pulsation. Our 2007--2011 data confirm V7
as a W~Vir variable, with a best-fit period of 13.9995~days. Figure~9 (top
panel) plots its phased light curve.

To the best of our knowledge, there is no previously published light curve for
V8; thus its classification has remained unclear, apart from the fact that it
lies near the tip of the red-giant branch in the HRD\null.  Our light curve,
shown in Figure~9 (bottom panel), shows that it is a semiregular variable, with
characteristic alternating depths of minima. Periodograms for the individual
annual seasons give periods ranging from 65.1 to 80.1~days; for the entire data
set, the analysis gives an average period of about 70.4~days. Several similar
variables are known, for example, in M13 (e.g., Osborn \& Fuenmayor 1977;
Kopacki et al.\ 2003)---which, as noted in \S2, has a very similar CMD to that
of M79.

{\bf V13 and V14:} V13 is an RRab variable, discovered by Amigo et al.\ (2011),
who give a period of 0.6906617~days. Kains et al.\ (2012) found a period of
0.689391~days. Our data from 2007 to 2011 are best fit with a very similar
period of 0.689399~days, as shown in the phased light curves in Figure~10 (top
panel). There is significant crowding at the location of V13, and it is located
near the edge of the 1.3m- field, making our data unusually noisy.

V14 is an RRc type variable that was discovered by Kains et al.\ (2012). Our
data confirm their period of 0.323733~days, as shown in Figure~10 (bottom panel).

The data for the five variable stars are included in this paper as a
machine-readable table. Table~5 illustrates the format of the table.

% 10/27/15: approx mags/colors for these variables:
% 
% Magnitude diffs rel to PAGB in Robin's file from 6 random lines:
% 
% B:
%                                                                 avg
% V6      4.161   4.18    3.749   4.107   3.934   3.602           3.9555                       
% V7      2.291   1.963   2.356   1.645   1.608   2.402		2.04416
% V8      1.698   1.676   1.762   1.817   1.931   2.037            1.82016                               
% V13     3.46    4.008   4.762   4.053   3.905   3.506		3.949
% V14     3.941   3.797   3.698   3.909   3.799   3.752           3.816		
%                 
% V:
%                                                                 avg
% V6      4.162   4.151   3.827   4.117   3.942   3.721		3.986666
% V7      1.784   1.636   1.875   1.361   1.238   1.905   	1.6331
% V8      .638    .606    .669    .676    .778    .871    	.706333
% V13     3.564   3.958   4.431   3.955   3.824   3.542   	3.879
% V14     3.969   3.878   3.768   4.003   3.846   3.862   	3.88766
% 
% Using for the PAGB B=12.48  V=12.20 we get:
% 
%         B       V       B-V
% V6      16.44   16.19	.25
% V7      14.52   13.83   .69
% V8      14.30   12.91	1.39
% V13     16.43   16.08	.35
% V14     16.30   16.09	.21

%----------------------------------------------------------------------------

\begin{figure}
\begin{center}
\includegraphics[width=2.8in]{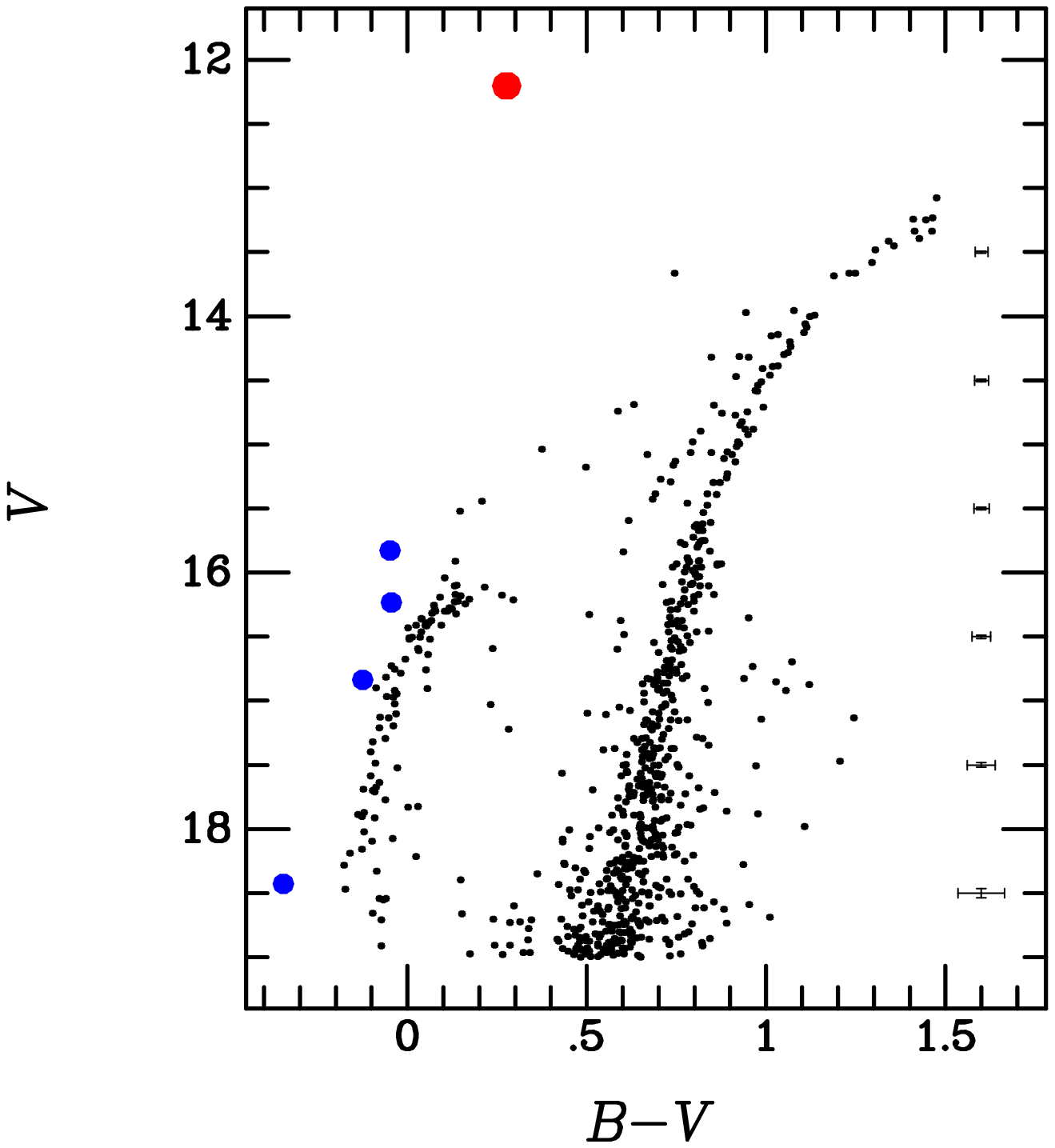}
\includegraphics[width=2.8in]{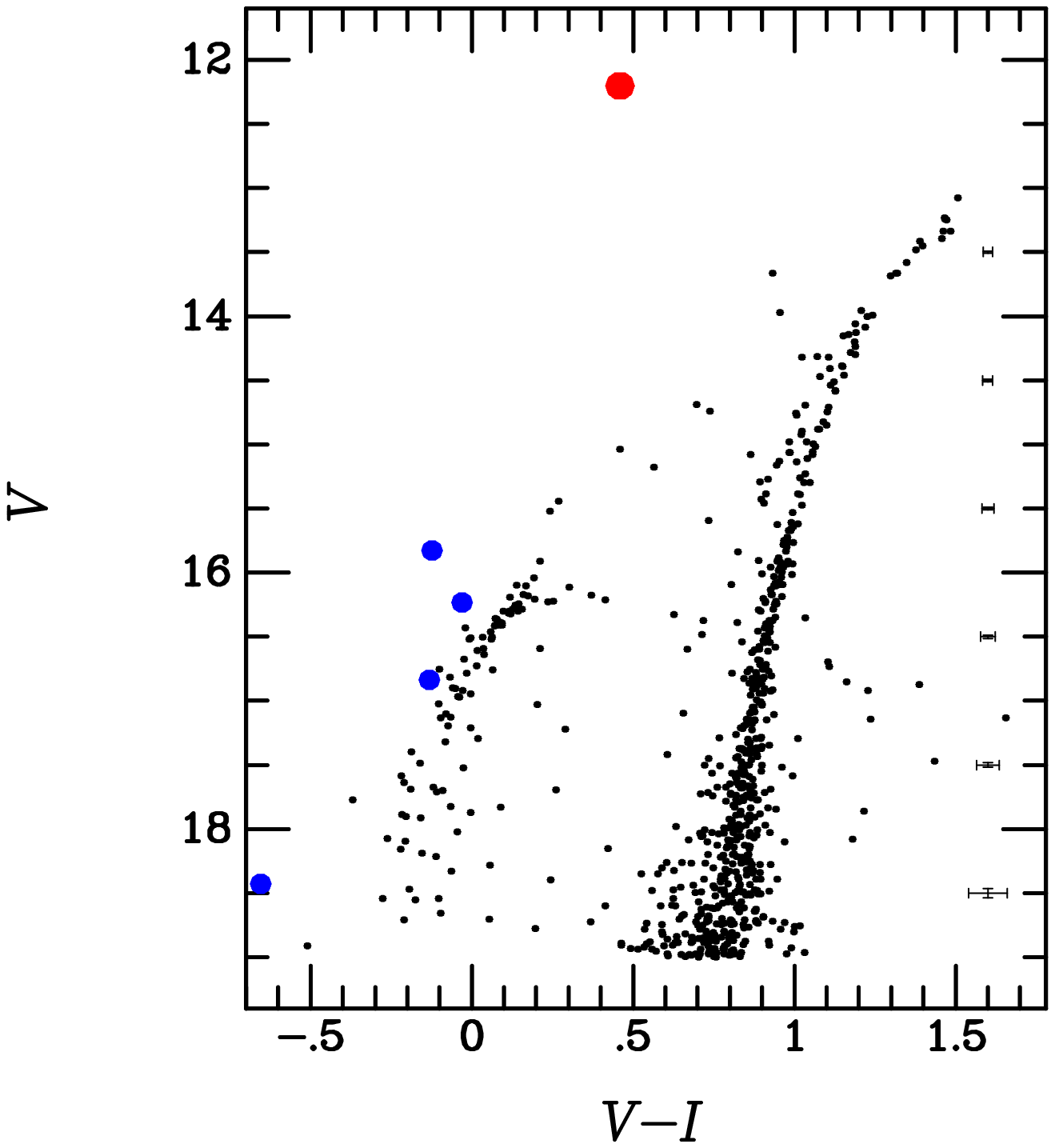}
\vskip0.2in
\includegraphics[width=4.5in]{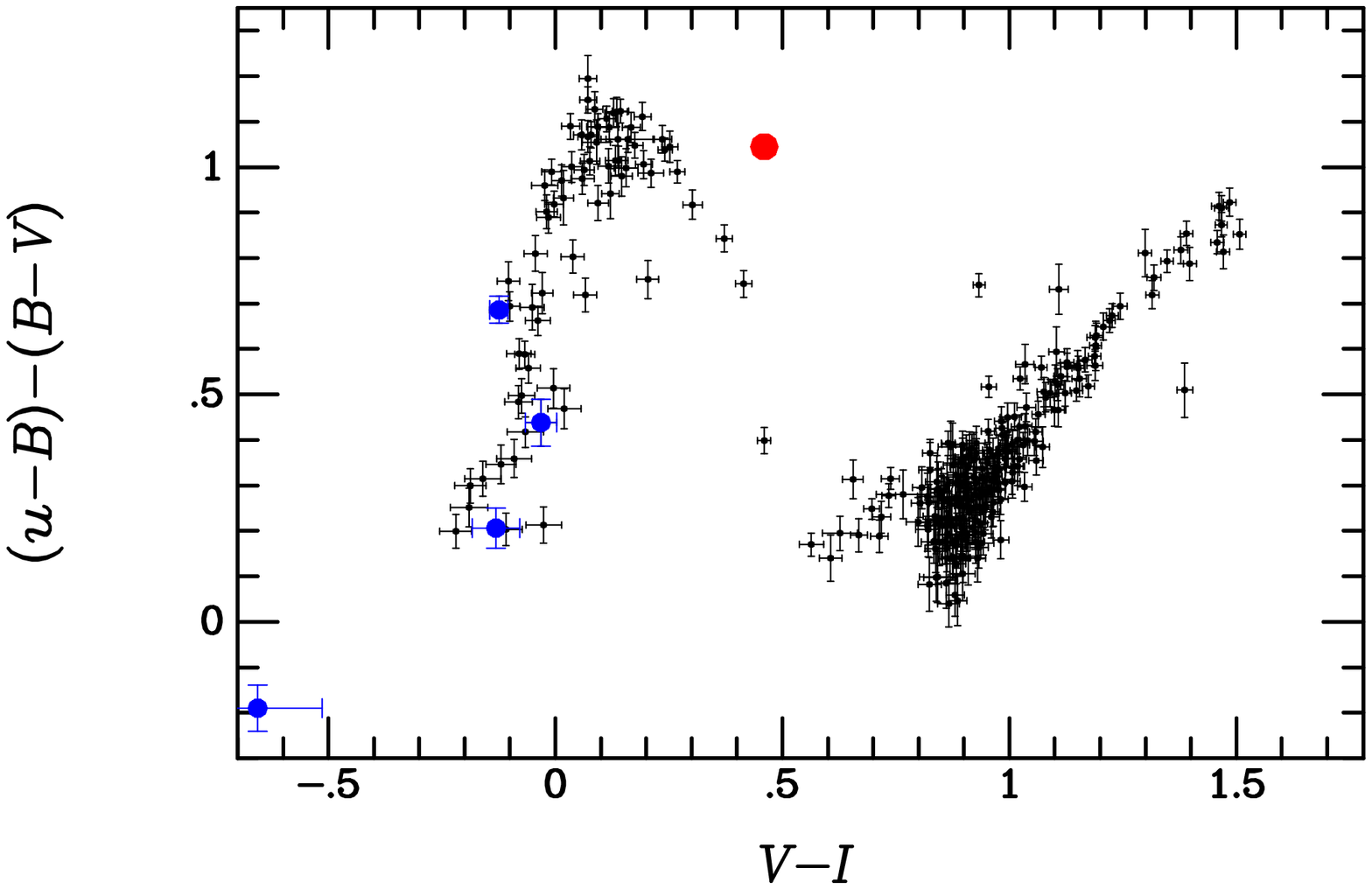}
\figcaption{\footnotesize
(a and b): Color-magnitude diagrams for M79 in $V,B-V$ and $V,V-I$\null. Stars
lying within $4\farcm8$ of the cluster center are plotted. Average photometric
error bars as a function of $V$ magnitude are plotted on the right-hand sides.
The {\it filled red circle\/} shows the location of the luminous PAGB star. Four
AGB-manqu\'e stars evolving off the horizontal branch are shown as {\it filled
blue circles\/}. (c): The gravity-sensitive $(u-B)-(B-V)$ color index plotted
against the temperature-sensitive $V-I$ color, for stars brighter than
$V=17.75$. The sequence peaking on the left-hand side of the diagram is the
horizontal branch. The sequence on the right-hand side is the subgiant and
red-giant branch. The PAGB star stands out because of its low $\log g$ and very
large Balmer jump. (The small clump at about $V-I\simeq0.65$ is foreground field
stars.)}
\end{center}
\end{figure}

\begin{figure}
\begin{center}
\plotone{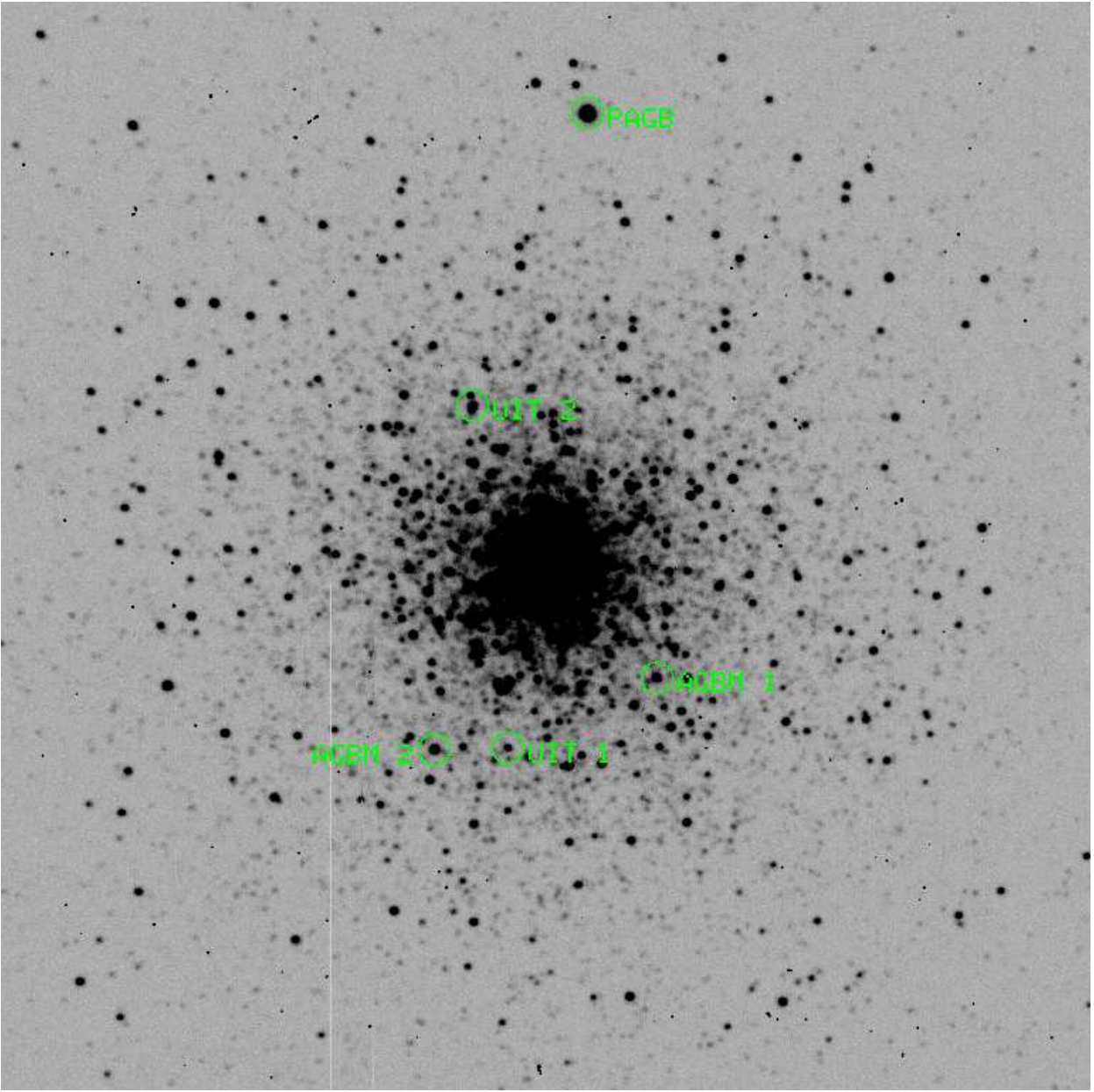}
\figcaption{
Finding chart for the PAGB and AGB-manqu\'e stars in M79 listed in Table~1, made
from our 400-s $u$-band exposure with the CTIO 0.9-m telescope. The field is
$5\farcm0\times5\farcm0$ and has north at the top, east on the left. UIT~2 is
blended with a close, cooler star.}
\end{center}
\end{figure}

\begin{figure}
\begin{center}
\plotone{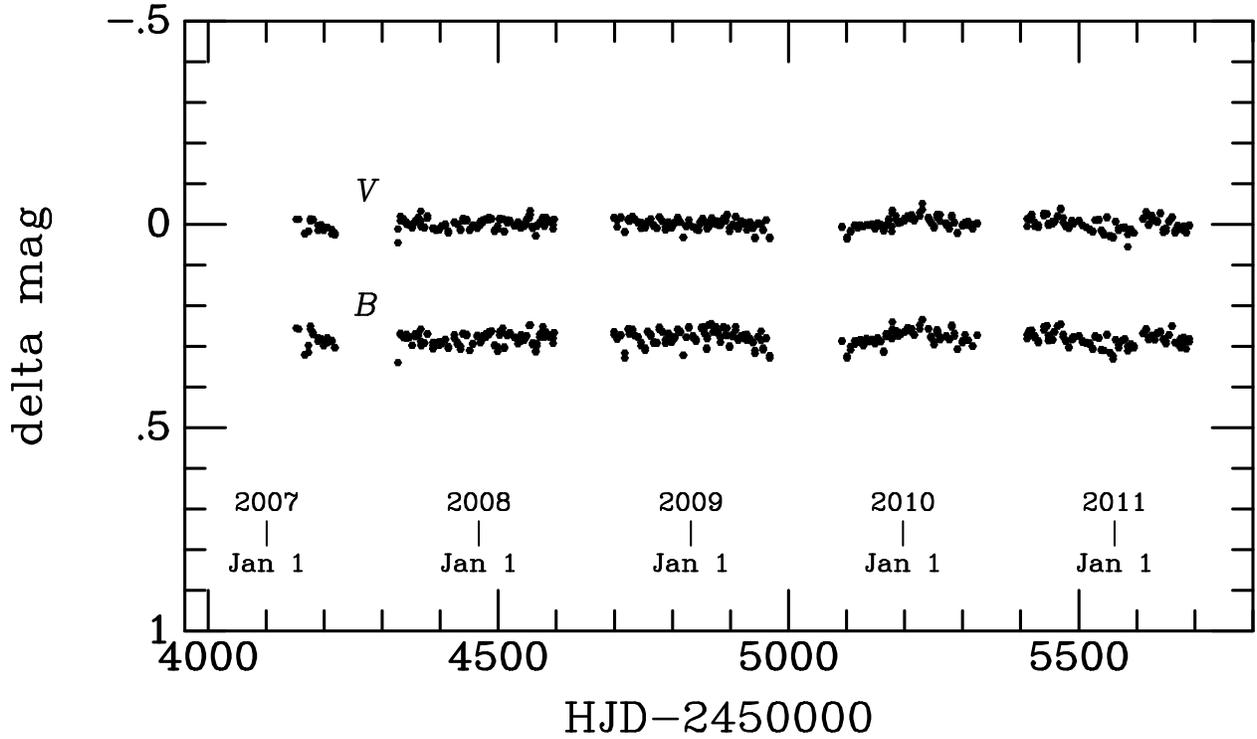}
\figcaption{
Differential photometry for the M79 PAGB star from early 2007 to mid-2011,
derived from observations with the SMARTS 1.3-m telescope and CCD camera. We
plot the magnitude difference between the PAGB star and the intensity sum of
eight nearby comparison stars, with arbitrary zero points. The star appears to
be constant within the errors.
}
\end{center}
\end{figure}

\begin{figure}
\begin{center}
\plotone{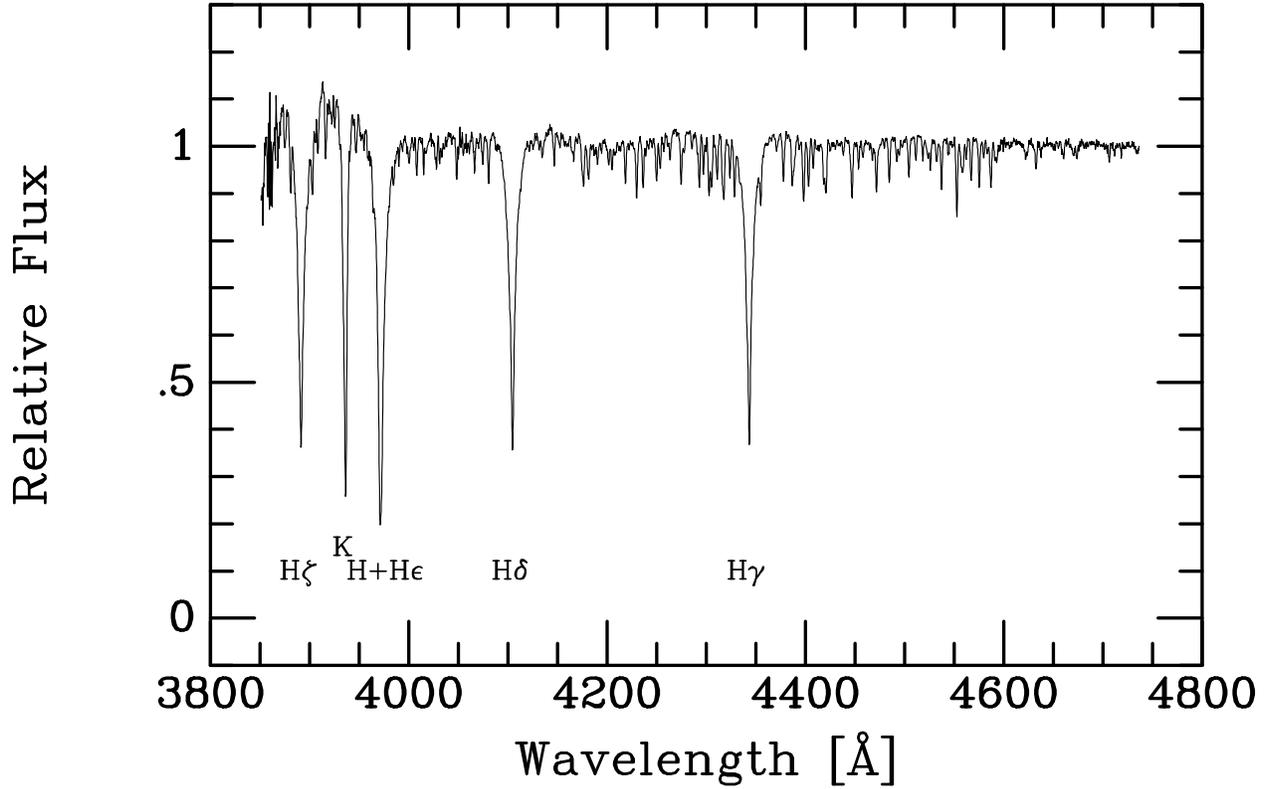}
\figcaption{
Mean spectrum of the M79 PAGB star (resolution $\sim$1.6~\AA), created by
combining data from 67 individual observations made with the SMARTS 1.5-m RC
spectrograph, and normalizing to a flat continuum. 
}
\end{center}
\end{figure}

\begin{figure}
\begin{center}
\plotone{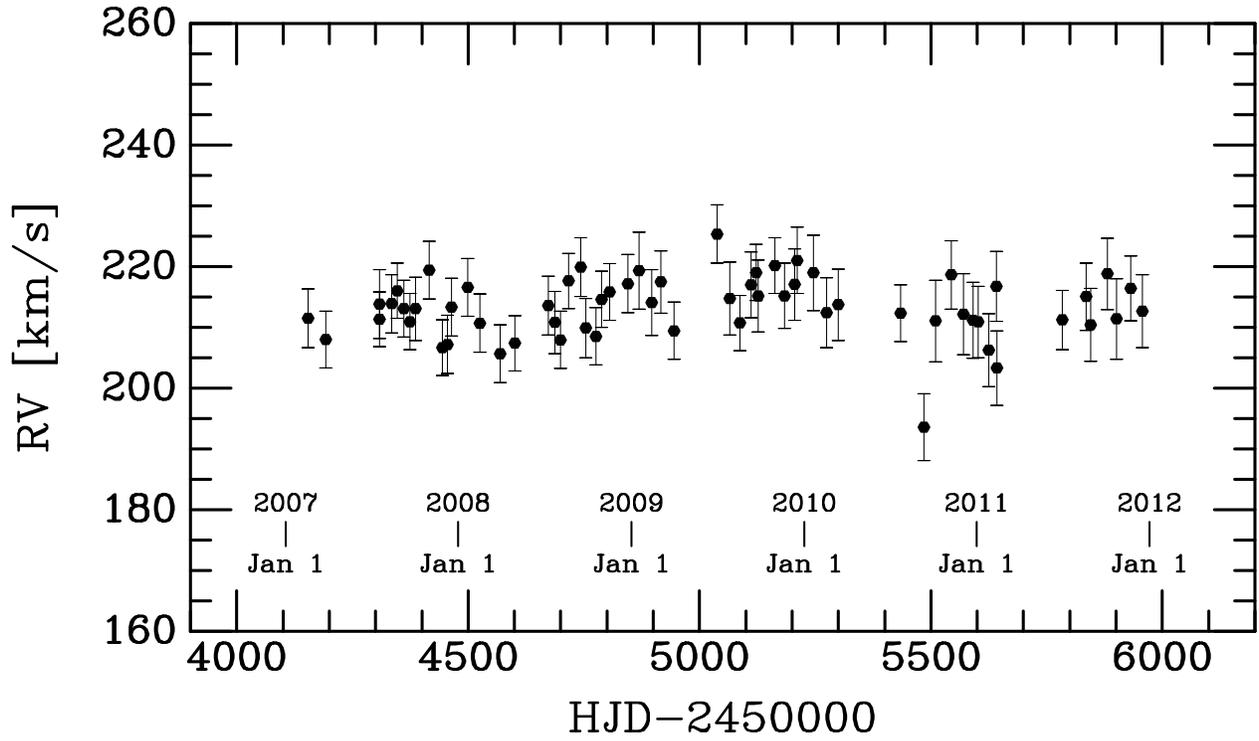}
\figcaption{
Radial-velocity measurements of the M79 PAGB star, based on observations made
with the SMARTS 1.5-m RC spectrograph. Within the errors, there is no evidence
for velocity variability. 
}
\end{center}
\end{figure}

\begin{figure}
\begin{center}
\plotone{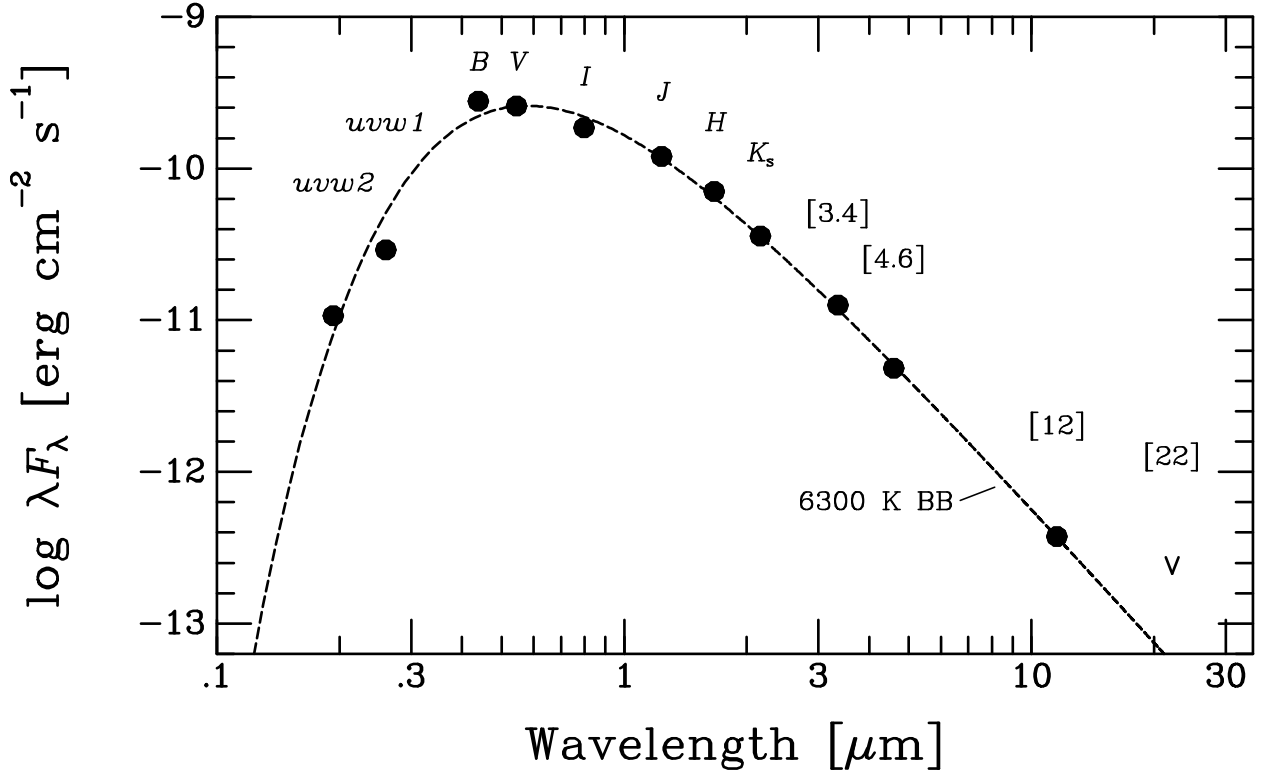}
\figcaption{
The SED of the M79 PAGB star, based on photometry by the {\it Swift\/} UVOT, our
ground-based photometry from Table~1, and the 2MASS and {\it WISE\/} surveys.
The point at $22\,\rm\mu m$ is an upper limit. See text for details. Also shown
is a blackbody spectrum for $\Teff=6300$~K, the temperature obtained in a
spectroscopic analysis of the star, normalized to the $V$ magnitude. There is no
evidence in the near- or mid-IR for circumstellar dust.
}
\end{center}
\end{figure}

\begin{figure}
\begin{center}
\includegraphics[height=4.5in]{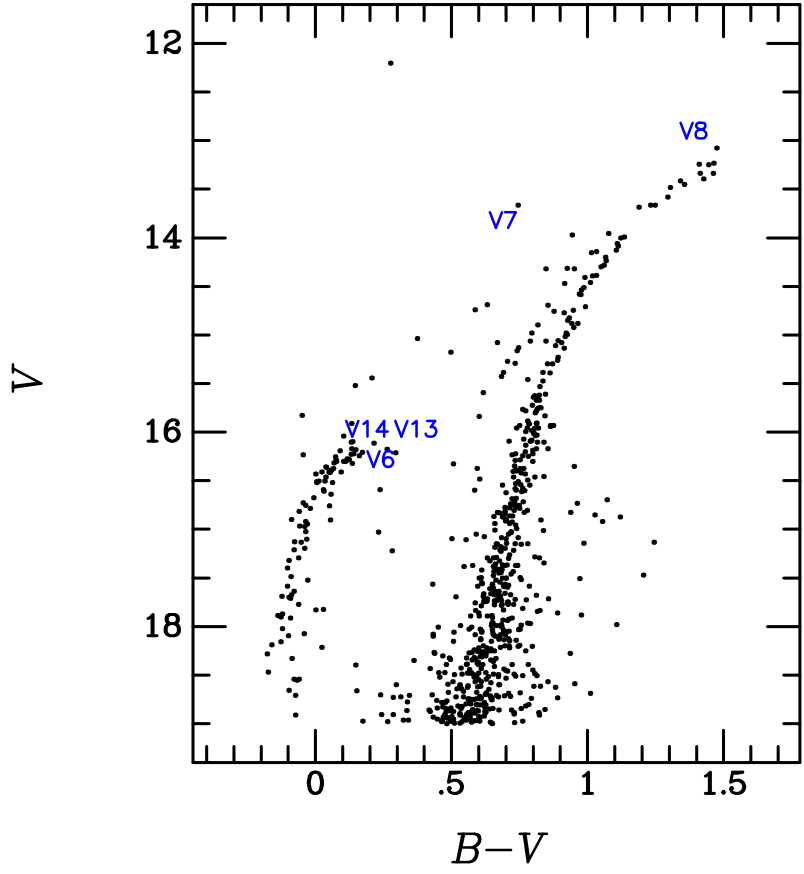}
\figcaption{
Color-magnitude diagram of M79 (taken from Figure~1a), with nominal locations of
five variable stars indicated by their designations.
}
\end{center}
\end{figure}

\begin{figure}
\begin{center}
\plotone{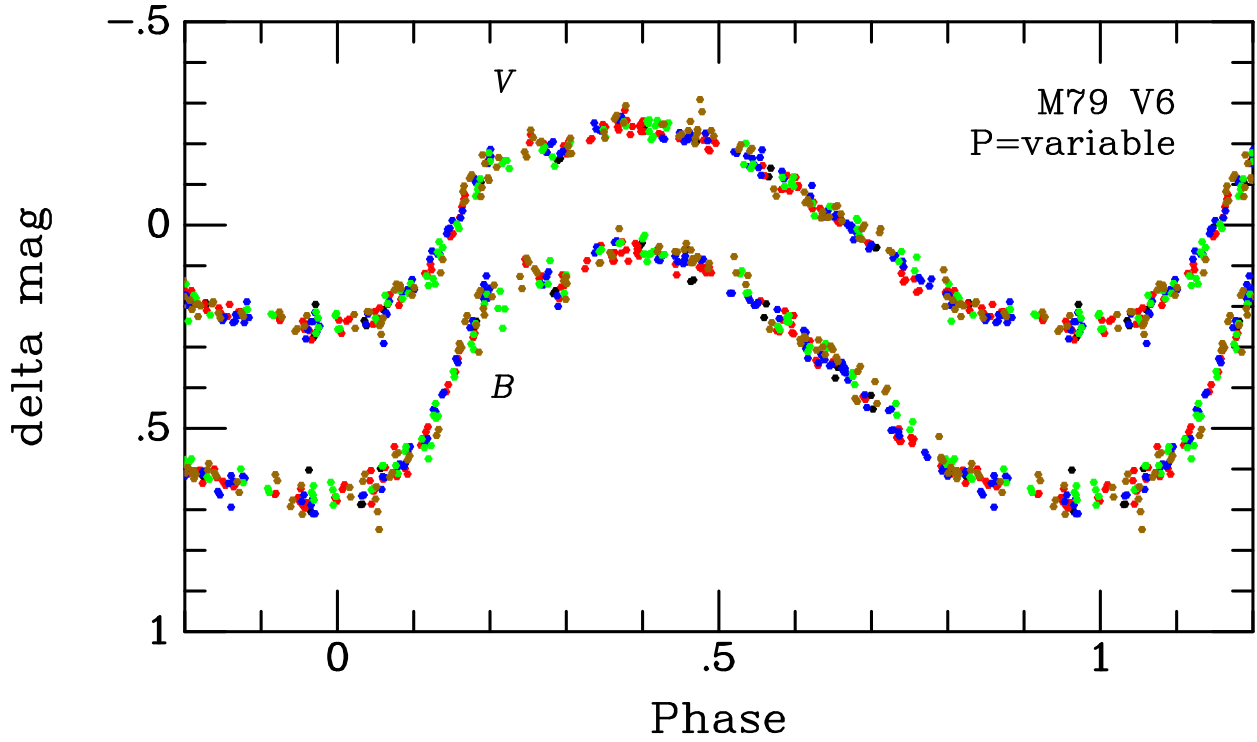}
\figcaption{
Light curves of the RRc variable M79-V6 in $V$ and $B$, phased according to the
seasonal elements given in Table~4. Color coding is: black (2007), red
(2007-2008), blue (2008-2009), green (2009-2010), and brown (2010-2011). The
light-curve shape appears constant, but the phases and periods are variable.
}
\end{center}
\end{figure}

\begin{figure}
\begin{center}
\includegraphics[height=3.4in]{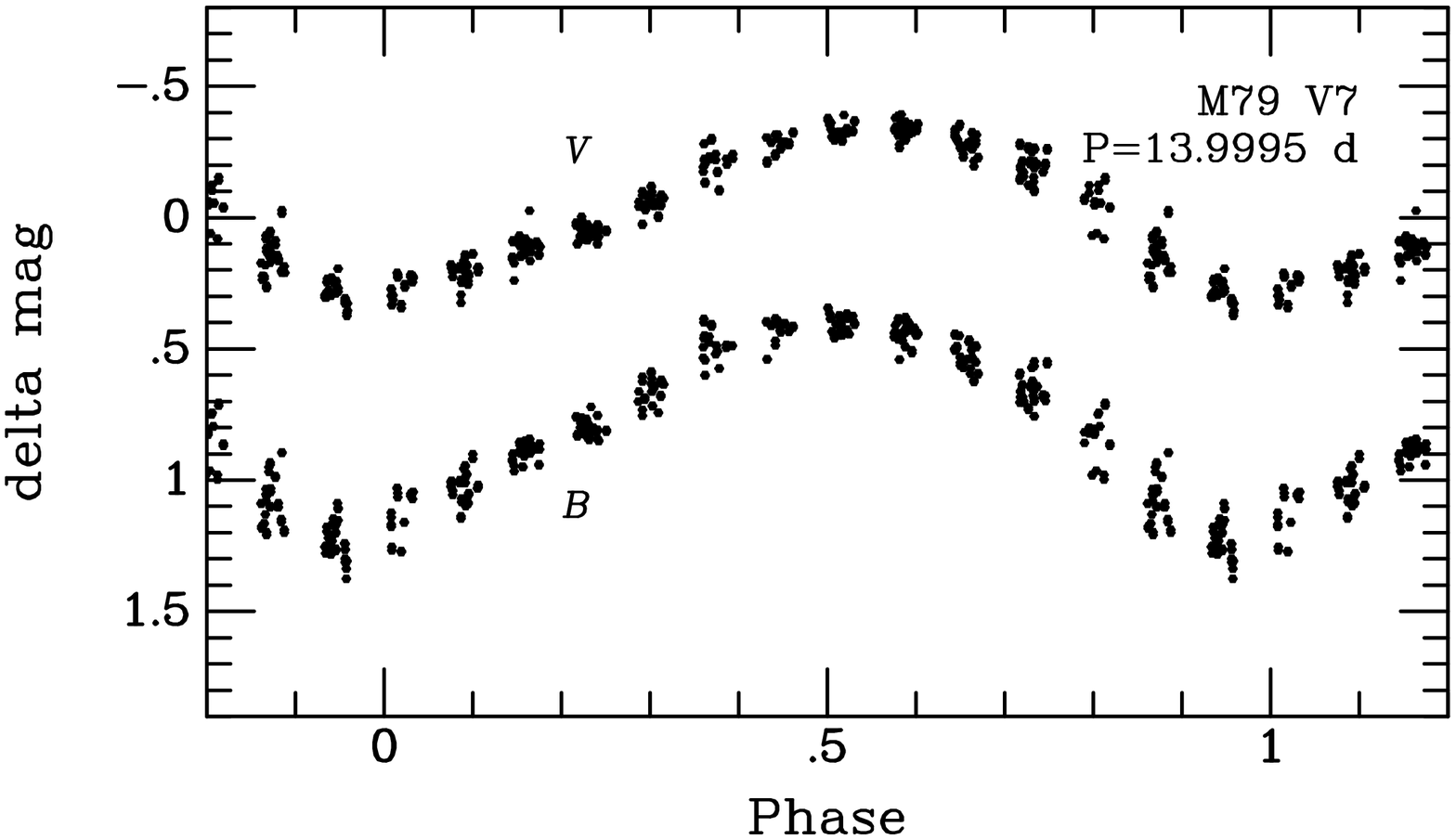}
\vskip0.2in
\includegraphics[height=3.4in]{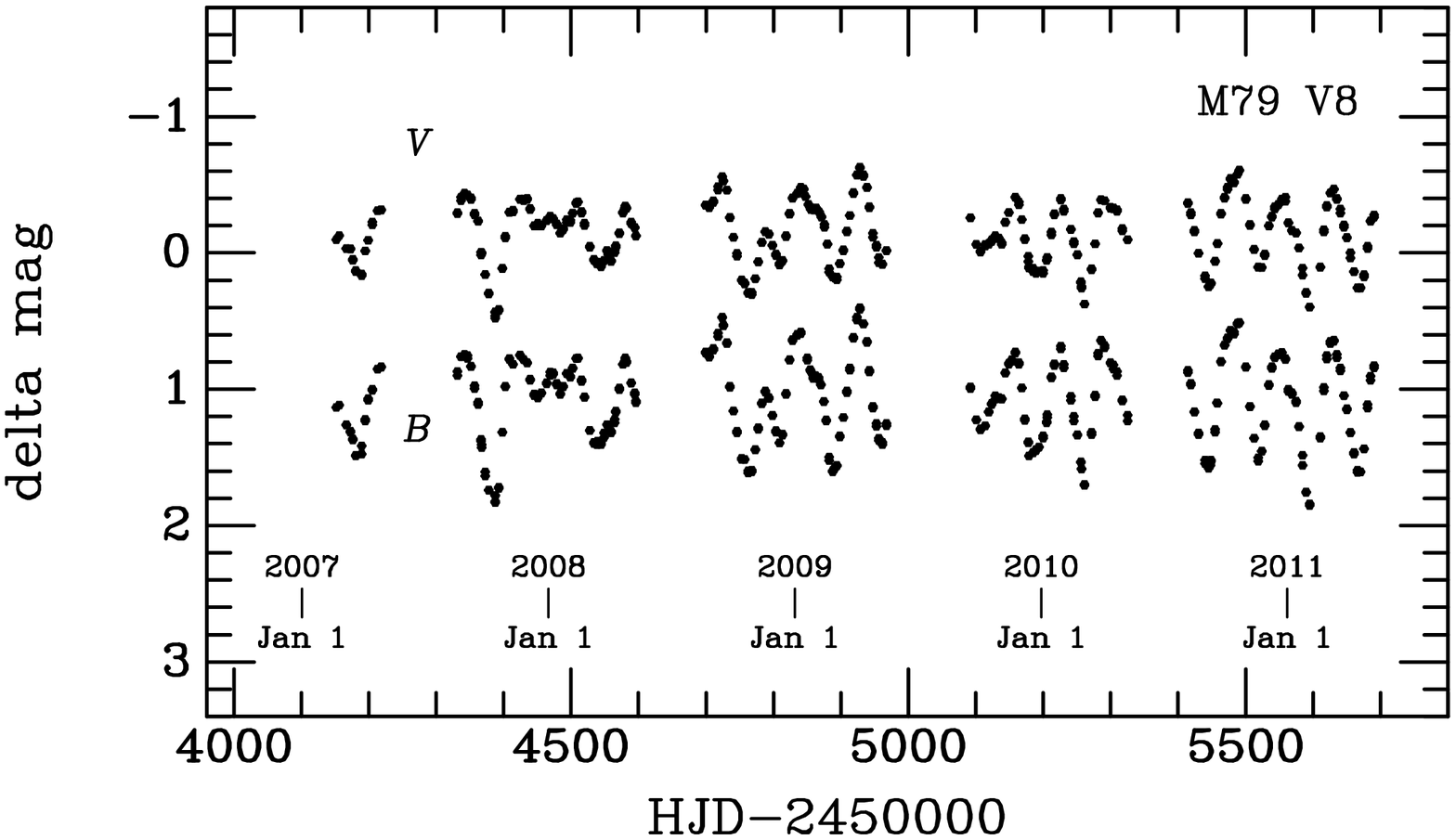}
\figcaption{
{\bf Top:} Light curves of the W~Virginis-type Cepheid M79-V7 in $V$ and $B$,
phased with a period of 13.9995~days. 
{\bf Bottom:} Light curves of the semiregular variable M79-V8 in $V$ and $B$.
The characteristic pulsation period varies from about 65 to 80~days.
In both plots, the magnitude zero-points are arbitrarily chosen. 
}
\end{center}
\end{figure}

\begin{figure}
\begin{center}
\includegraphics[height=3.4in]{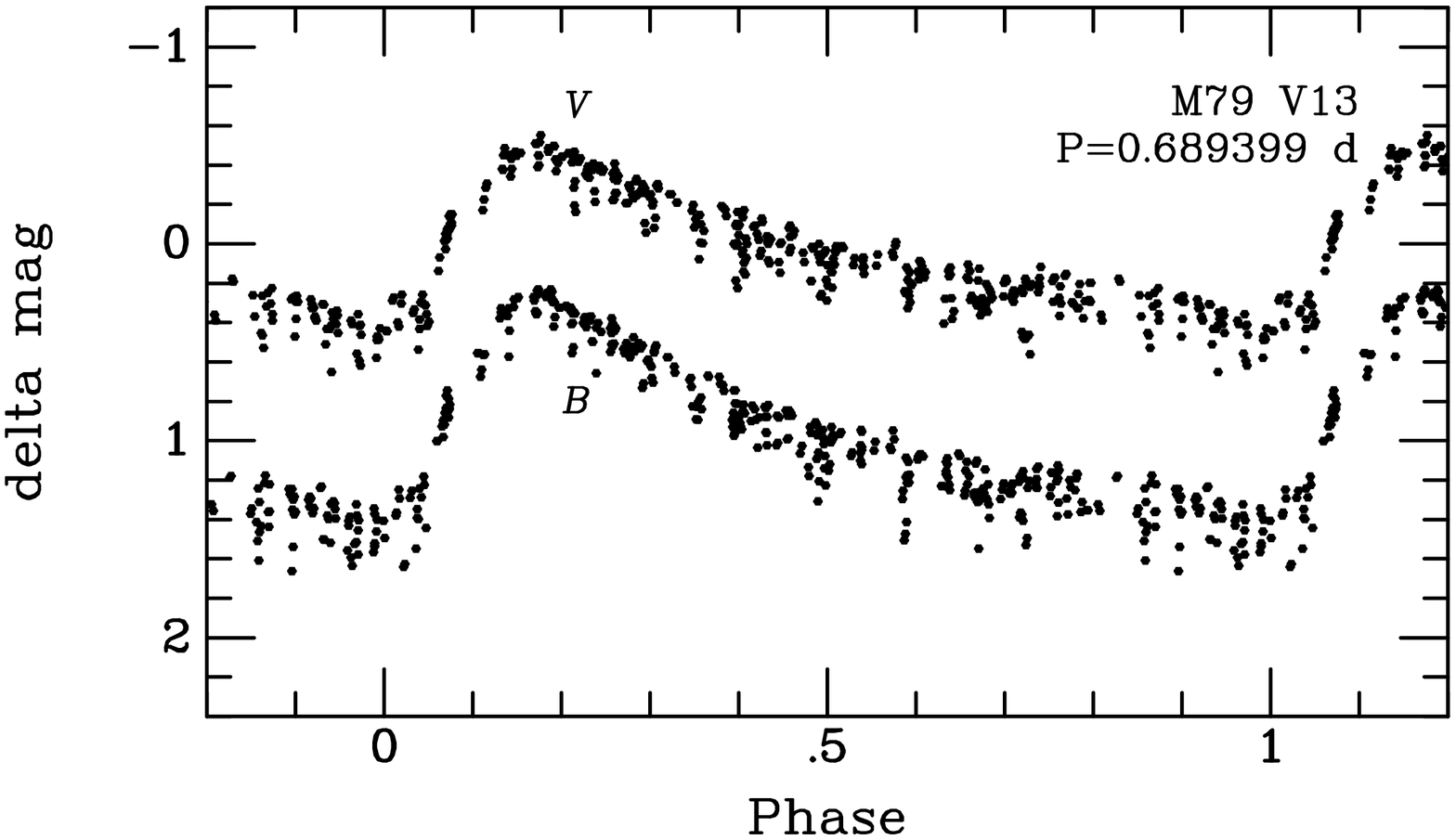}
\vskip0.2in
\includegraphics[height=3.4in]{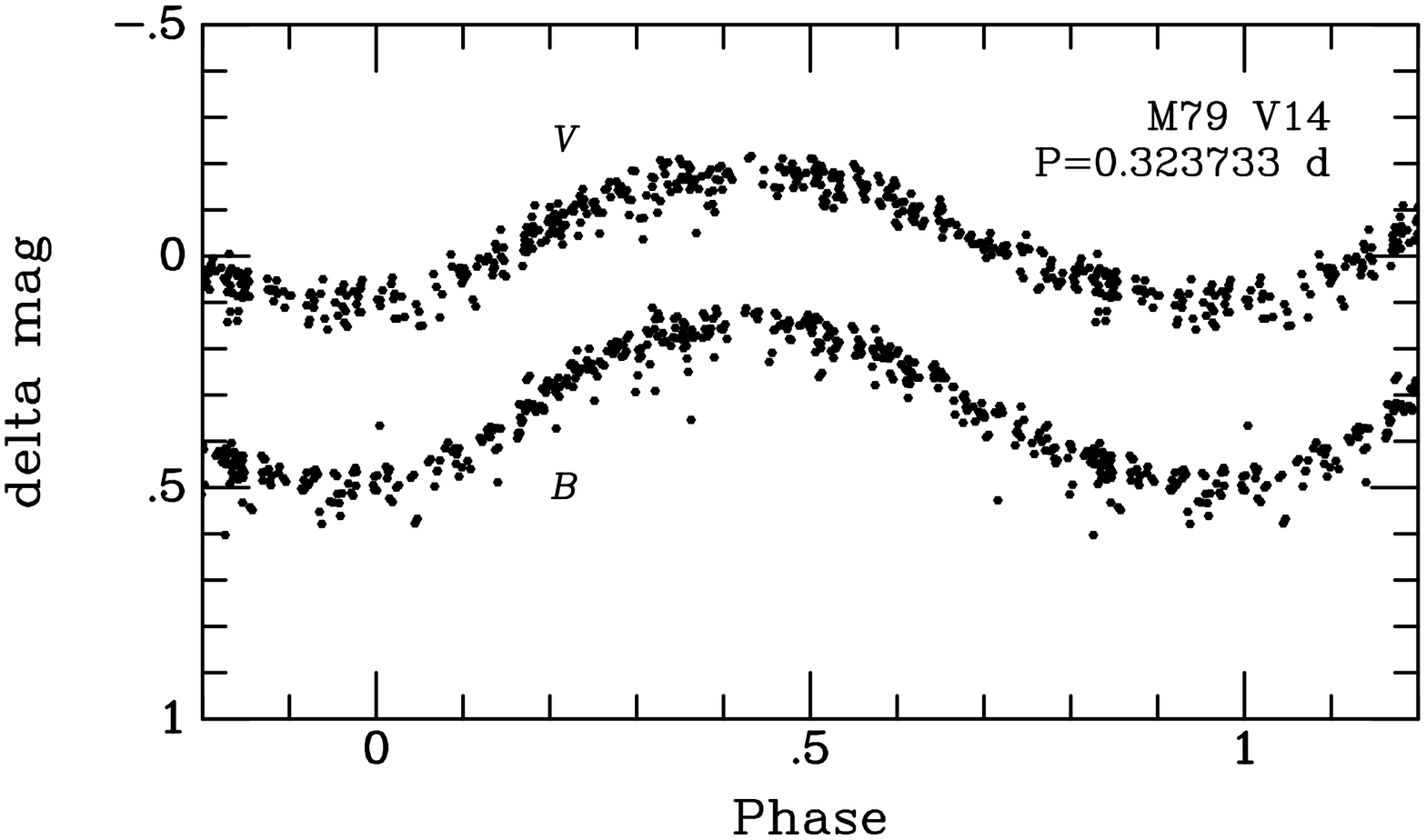}
\figcaption{
{\bf Top:} Light curves of the RRab variable M79-V13 in $V$ and $B$, phased
according to the elements given in Table~3. 
{\bf Bottom:} Light curves of the RRc variable M79-V14 in $V$ and $B$, phased
according to the elements given in Table~3. 
}
\end{center}
\end{figure}

\clearpage

\begin{deluxetable}{lllllll}
\tablewidth{0 pt}
\tablecaption{PAGB and AGB-M Stars in M79}
% \tabletypesize{\footnotesize}
\tabletypesize{\scriptsize}
\tablehead{
\colhead{Designation} &
\colhead{RA (J2000)} &
\colhead{Dec (J2000)} &
\colhead{$V$} &
\colhead{$B-V$} &
\colhead{$V-I$} &
% \colhead{$u-B$} &
\colhead{$(u-B)-(B-V)$} 
}
\startdata
PAGB\tablenotemark{a} & 5:24:10.36 & $-24$:29:20.6 & $12.203\pm0.008$  & 
   $ 0.277\pm0.013 $ & $ 0.460\pm0.013$ & $ 1.045\pm0.018$ \\
AGBM 1 & 5:24:08.86 & $-24$:32:00.1 & $16.234\pm0.021$ & $-0.046\pm0.038 $ & 
   $-0.031\pm0.034$ & $ 0.438\pm0.051$ \\
AGBM 2 & 5:24:13.45 & $-24$:32:20.7 & $15.827\pm0.012$ & $-0.048\pm0.021 $ & 
   $-0.124\pm0.020$ & $ 0.686\pm0.030$ \\
UIT 1\tablenotemark{b}& 5:24:11.93 & $-24$:32:20.1 & $18.427\pm0.030$  & 
   $-0.346\pm0.039 $ & $-0.656\pm0.141$ & $-0.190\pm0.051$ \\
UIT 2  & 5:24:12.69 & $-24$:30:43.4 & $16.834\pm0.020$ & $-0.125\pm0.031 $ & 
   $-0.131\pm0.052$ & $ 0.206\pm0.044$ \\
\enddata
\tablenotetext{a}{The PAGB star is also cataloged as TYC 6479-422-1 and 2MASS
J05241036$-$2429206.}
\tablenotetext{b}{UIT~1 is very faint in our frames, except in $u$, and the data
are only approximate.}
\end{deluxetable}

\begin{deluxetable}{lllccccc}
\tablewidth{0 pt}
\tablecaption{Absolute Magnitudes of PAGB Candidates in Globular Clusters}
% \tabletypesize{\footnotesize}
% \tabletypesize{\scriptsize}
\tablehead{
\colhead{} &
\colhead{} &
\colhead{} &
\colhead{Photometry} &
\colhead{} &
\colhead{Distance} &
\colhead{} \\
\colhead{Cluster} &
\colhead{Star} &
\colhead{$V$} &
\colhead{Source} &
\colhead{$(m-M)_V$} &
\colhead{Source} &
\colhead{$M_{V,0}$} 
}
\startdata
$\omega$~Cen &  ROA 24 &  10.89 & (1)  &    13.99  & (4)  & $-3.10$ \\
NGC 5986     &  PAGB 1 &  12.65 & (2)  &    16.05  & (2)  & $-3.40$ \\
\qquad$''$   &  PAGB 2 &  12.76 & $''$ &    $''$   & $''$ & $-3.29$ \\
M79          &  PAGB   &  12.20 & (3)  &    15.66  & (5)  & $-3.46$ \\
\enddata
\tablerefs{References for photometry of the PAGB stars and
apparent $V$-band distance moduli of clusters are: 
(1)~van Leeuwen et al.\ (2000); (2)~Alves et al.\ (2001); (3)~this paper;
(4) Matsunaga et al.\ (2006); (5)~Kains et al.\ (2012).} 
\end{deluxetable}

% $E(B-V)$
% 
% 0.12	
% 0.28	
% 
% 0.01	

\begin{deluxetable}{llcc}
\tablewidth{0 pt}
\tablecaption{M79 Variables in our 1.3-m Monitoring Field}
\tablehead{
\colhead{Designation} &
\colhead{Type} &
\colhead{$T_0$ [HJD$-$2450000]} &
\colhead{$P$ [days]}  
}
\startdata
V6  & RRc & var.\tablenotemark{a} & var.\tablenotemark{a} \\
V7  & CW  & 4139.450		  & 13.9995		  \\
V8  & SRd & $\dots$		  & 65.1--80.1  	  \\
V13 & RRab & 4151.442             & 0.689399              \\
V14 & RRc & 4151.207		  & 0.323733		  \\
\enddata
\tablenotetext{a}{See Table 4 for seasonal elements.}
\end{deluxetable}

\begin{deluxetable}{lcc}
\tablewidth{0 pt}
\tablecaption{Elements for M79-V6}
\tablehead{
\colhead{Season} &
\colhead{$T_0$ [HJD$-$2450000]} &
\colhead{$P$ [days]}  
}
\startdata
2007       & 4151.216 & 0.339122 \\
2007--2008 & 4151.244 & 0.339076 \\
2008--2009 & 4151.295 & 0.339036 \\
2009--2010 & 4151.253 & 0.339051 \\
2010--2011 & 4151.253 & 0.339145 \\
\enddata
\end{deluxetable}

\begin{deluxetable}{lc}
\tablewidth{0 pt}
\tablecaption{Differential Photometry of Variable Stars in M79\tablenotemark{a}}
\tablehead{
\colhead{HJD$-$2450000} &
\colhead{Delta Magnitude\tablenotemark{b}}  
}
\startdata
\multispan2{\hfil M79 V6 $B$ Magnitudes \hfil} \\
4151.57438 &   3.7985 \\
4156.58235 &   3.8212 \\
4166.53713 &   3.4525 \\
4166.53799 &   3.4358 \\
4172.51951 &   3.8042 \\
\enddata
\tablenotetext{a}{Table 5 is published in its entirety in the electronic 
edition of the journal. A portion is shown here for guidance regarding its form
and content.}
\tablenotetext{b}{The magnitudes are differential with respect to the sum of the
intensities of eight comparison stars in the 1.3-m field.}
\end{deluxetable}

\end{document}